\documentclass[pre,twocolumn,twoside,showpacs,floatfix]{revtex4}
\usepackage{epsf,times,dcolumn}

\newcommand{\knn}{{\langle k_{\rm nn} \rangle}}
\newcommand{\gnn}{{\langle g_{\rm nn} \rangle}}
\begin{document}
\title{Betweenness centrality correlation in social networks} 
\author{K.-I. Goh, E. Oh, B. Kahng, and D. Kim \\}
\affiliation{
\mbox{School of Physics and Center for Theoretical Physics,
Seoul National University, Seoul 151-747, Korea}}
\date{\today}
\begin{abstract}
Scale-free (SF) networks exhibiting a power-law degree distribution can 
be grouped into the assortative, dissortative and neutral networks 
according to the behavior of the degree-degree correlation coefficient.  
Here we investigate the betweenness centrality (BC) correlation for 
each type of SF networks. While
the BC-BC correlation coefficients behave similarly to the degree-degree
correlation coefficients for the dissortative and neutral networks, 
the BC correlation is nontrivial for the assortative ones found
mainly in social networks. 
The mean BC of neighbors of a vertex with BC $g_i$ is almost 
independent of $g_i$, implying that each person is surrounded by 
almost the same influential environments of people no matter how 
influential the person is.  
\end{abstract}
\maketitle
Recently there have been considerable efforts to understand 
complex systems in terms of random graph, consisting of vertices 
and edges, where vertices (edges) represent constituents 
(their interactions)~\cite{review1,review2}. 
An interesting feature emerging in such complex networks is 
a power-law degree distribution, $p_d(k)\sim k^{-\gamma}$, 
where the degree $k$ is the number of edges incident upon 
a given vertex~\cite{www,huberman}. 
Networks displaying the power-law degree 
distribution are called the scale-free (SF) networks. 
Barab\'asi and Albert (BA) introduced an evolving model 
illustrating such a SF network~\cite{ba}. 
The degree distribution of the BA model follows a power-law 
$p_d(k)\sim k^{-3}$. 
While the BA-type models are meaningful as the first step to generate 
SF network, they are too simple to accord with real-world 
networks, exhibiting nontrivial degree-degree and other correlations.\\ 

SF networks can be grouped into three types 
according to the behavior of the degree-degree correlation 
coefficient~\cite{newman,newman2}. 
They are the ones exhibiting the assortative, 
dissortative, and neutral mixing on their degree. 
For the network of the assortative (dissortative) mixing, 
called the assortative (dissortative) network,  
a vertex with large degree tends to connect to vertices 
with large (small) degree, while for the network of the neutral mixing, 
there is no such tendency. The assortative network  
can be found in social networks such as the coauthorship network, 
the actor network and so on, and the dissortative network in information 
networks such as the Internet and the World-wide Web, and in biological 
networks such as protein interaction networks and neural networks.
While such assortative and dissortative networks appear in real world, 
the neutral network, $i.e.$, the network of the neutral mixing on their 
degree, appears in in silico networks such as the BA model and 
the copying model \cite{copying} with $\gamma=3$.\\ 
 
While degree is a fundamental quantity describing the topology of the 
SF network, it was shown that the betweenness centrality (BC) is 
another important quantity to characterize how influential a vertex 
is in communications between each pair of vertices~\cite{freeman,newman1}. 
To be specific, let us suppose communication paths between a pair 
of vertices $(i,j)$ are the shortest pathways 
and let the number of such pathways denoted by $c(i,j)$. 
Among them, the number of the shortest pathways running 
through a vertex $k$ is denoted by $c_k(i,j)$ and the fraction 
by $g_k(i,j)=c_k(i,j)/c(i,j)$.
Then the BC of the vertex $k$ is defined 
as the accumulated amount of $g_k(i,j)$ over all pairs, $i.e.$, 
$g_k=\sum_{\{(i.j)\}}{c_k(i,j)}/{c(i,j)}=\sum_{\{(i,j)\}}g_k(i,j)$.
The BC distribution follows a power law for SF networks, 
$p_{b}(g) \sim g^{-\eta}$, 
where $g$ means BC and the exponent $\eta$ turns out to be 
robust as either $\eta \approx 2.2$ or $\eta =2.0$, 
independent of the degree exponent as long as 
$2 < \gamma \le 3$~\cite{load,class}.\\

For the BA-type model, it was shown that the BC is related 
to the degree via the relation~\cite{load} 
\begin{equation}
g\sim k^{(\gamma-1)/(\eta-1)}. 
\label{g_k_relation}
\end{equation} 
Thus the vertices with larger degree are much more influential 
to others in communications. 
Due to this relation, one may think that the BC-BC correlation 
would behave similarly to the degree-degree correlation. 
In this paper, we report that while for the dissortative 
and neutral network, the BC-BC correlation coefficients behave 
similarly to the degree-degree correlation coefficients, 
for the assortative network, 
the degree-BC relation Eq.~(\ref{g_k_relation}) is nontrivial, 
leading to that the BC-BC correlation is very weakly assortative, 
i.e., the mean BC of neighbors of a certain vertex with BC $g_i$ is 
almost independent of $g_i$.\\
 
The degree-degree correlation~\cite{redner,serguei} was investigated 
in terms of the correlation function between the remaining 
degrees of the two vertices on each side of an edge, 
where the remaining degree means the degree of that vertex minus 
one \cite{newman}. 
First one defines the joint probability $e_{d}(j,k)$ that 
the two vertices on each side of a randomly chosen link have $j$ and $k$ remaining degrees, respectively.
Then the normalized correlation coefficient is 
defined as 
\begin{equation}
r_{d}={1\over \sigma_{d}(q)^2}\sum_{j,k}jk 
\{e_{d}(j,k)-q_{d}(j) q_{d}(k)\},
\label{r_d}
\end{equation} 
where $q_{d}(k)$ is the normalized distribution of the 
remaining degree $q_{d}(k)=(k+1)p_{d}(k+1)/\sum_j 
j p_{d}(j)$, 
and $\sigma_{d}(q)^2=\sum_k k^2 q_d(k) -[\sum_k kq_d(k)]^2$. 
Recently Newman called this quantity the degree assortativity 
coefficient~\cite{newman2}.
For the assortative (dissortative) networks, $r_d$ is 
positive (negative), and for the neutral networks, $r_d=0$.
On the other hand, the degree-degree correlation was also 
investigated in terms of the mean degree of neighbors of a 
vertex with degree $k$, denoted by $\knn(k)$~\cite{vespignani}. 
For the assortative (dissortative) networks, $\knn(k)$ increases 
(decreases) with increasing $k$, while the neutral networks, 
$\knn(k)$ is independent of $k$.\\ 

To study the BC-BC correlation, we introduce the BC-BC correlation 
coefficient, called the BC assortativity coefficient, in analogy with 
Eq.(\ref{r_d}) as 
\begin{equation}
r_{b}={1\over \sigma_{b}(q)^2}\sum_{\ell,m}\ell m 
\{e_{b}(\ell, m)-p_{b}(\ell) p_{b}(m)\},
\label{r_b}
\end{equation} 
where $e_{b}(\ell,m)$ is the joint probability that 
the BCs of the two vertices of a link are $\ell$ and $m$ 
and $\sigma_{b}(q)^2=\sum_{\ell} \ell^2 p_b(\ell) -
[\sum_{\ell} \ell p_b(\ell)]^2$. 
Moreover, similarly to $\knn$, we define the mean BC 
of neighbors of a vertex with BC $g$, denoted by $\gnn(g)$, 
through which we can check if the BC-BC correlation is assortative or 
dissortative. \\
 
\begin{table}[t]
\begin{tabular}{clcccccc}
\hline\hline
 Type& &Name & $N$ & $\langle k\rangle$ & $r_d$ & $r_b$ & Ref.\ \\
\hline
       &\vline& Videomovie & 29824 & 33.7 & 0.22 & 0.024 & \cite{imdb}\\
Actor  &\vline& TVminiseries & 33980 & 73.0 & 0.38 & 0.033 & \cite{imdb}\\
       &\vline& TVcablemovies & 117655 & 55.5 & 0.14 & 0.035 & \cite{imdb}\\
       &\vline& TVseries  & 79663 & 118.4 & 0.53 & 0.013 & \cite{imdb}\\
\hline
       &\vline& Neuroscience & 205202 & 11.8 & 0.60 & 0.057 & \cite{neuro}\\
Coauthor &\vline& Mathematics & 78835 & 5.50 & 0.59 & 0.091 & \cite{neuro}\\
       &\vline& cond-mat & 16264 & 5.85 & 0.18 & 0.086 & \cite{newman_pnas}\\
       &\vline& arXiv.org & 52909 & 9.27 & 0.36 & 0.057 & \cite{newman_pnas}\\
\hline\hline
\end{tabular}
\caption{Size $N$, mean degree $\langle k \rangle$, degree assortativity 
coefficient $r_d$, BC assortativity coefficient $r_b$ for a number of 
social networks.}
\end{table}

\begin{figure*}
\centerline{\epsfxsize=13.25cm \epsfbox{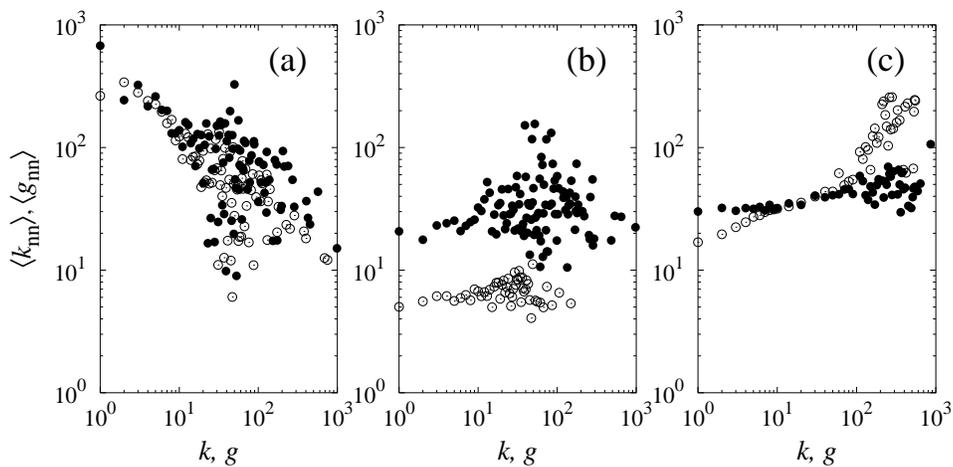}} 
\caption{
Plot of $\knn(k)$ ($\circ$) and $\gnn(g)$ ($\bullet$) 
for (a) the Internet on the level of autonomous systems (dissortative),
(b) the non-degenerate configuration model with $\gamma=3$ (neutral),
and (c) the coauthorship network in the field of neuroscience 
(assortative). All data are obtained from a single configuration.
}
\label{fig1}
\end{figure*}

We first check the BC-BC correlation for the network of 
the Internet on the level of autonomous systems as of January 2000 \cite{as}
and the so-called non-degenerate configuration model 
with $\gamma=3$ \cite{config,serguei,definition},
which belong to the dissortative and the neutral network, 
respectively. For these networks, $r_{b}$ is $-0.16$ ($<0$) 
and 0.02, respectively, which is close to their $r_d$ values of
$-0.18$ and $0.01$, respectively. 
Moreover, $\gnn(g)$ behaves similarly to $\knn(k)$ 
as shown in Fig.1a and b. However for the  
assortative networks, the coauthorship network for example,
$r_b$ is considerably smaller than $r_d$ often by one order of 
magnitude and is close to zero. The comparison of $r_b$ and 
$r_d$ for various social networks are tabulated in Table 1.  
The mean BC $\gnn(g)$ of neighbors of a vertex with BC $g$ 
increases with increasing $g$, however, the increasing rate 
is very low compared with that of $\knn(k)$, $i.e.$, it depends on $g$ 
very weakly (Fig.1(c)). Such a behavior appears in other social networks too. 
Since BC is regarded as a good measure of centrality, it implies
that the mean influence of neighbors of a person 
is almost the same regardless of the influence of the centered 
person. So a person is surrounded by almost the same 
influential people on average no matter how influential 
the centered person is, although a person who acquaints 
many people is likely to connect to people who also 
acquaint many others.\\

\begin{figure*}
\centerline{\epsfxsize=13.25cm \epsfbox{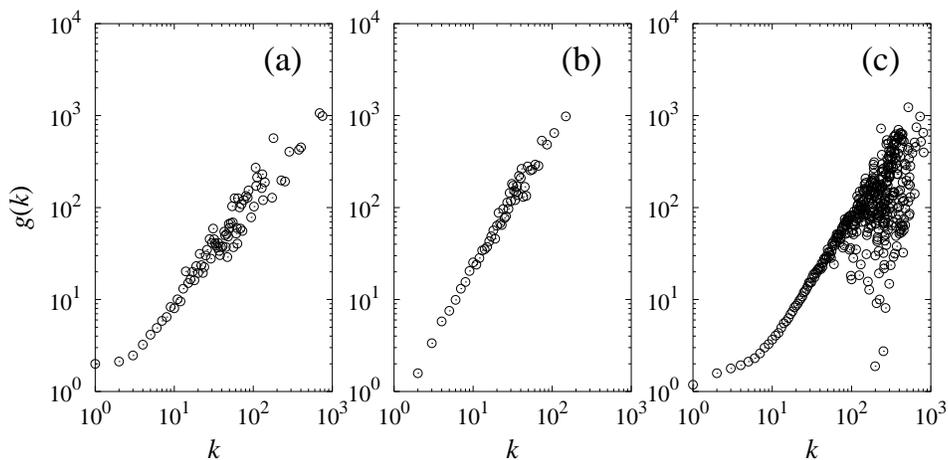}} 
\caption{Plot of the degree-BC relation 
for (a) the Internet on the level of autonomous systems (dissortative),
(b) the non-degenerate configuration model with $\gamma=3$ (neutral),
and (c) the coauthorship network in the field of 
neuroscience (assortative).}  
\label{fig2}
\end{figure*}

To understand the abnormal behavior of the BC-BC correlation 
in detail, we examine the degree-BC relation. 
In Fig.2, we compare the degree-BC relation $g(k)$ for the three types. 
While the relation of Eq.(\ref{g_k_relation}) holds
for the dissortative and the neutral networks,  
it breaks down for large $k$ for the assortative networks. 
Rather the BCs of large $k$ vertices cover wide range of values.
Since the vertices with large degree 
are located next to each other in the assortative network, 
the shortest pathways between a certain pair of vertices do not 
necessarily pass through such nearby hubs at the same time. Thus the BCs of 
the vertices with large $k$ fluctuate and the degree-BC 
correlation is nontrivial.
Next, we compare the clustering coefficient $C(k)$ as a 
function of degree $k$ with that $C(g)$ as a function of BC $g$ in Fig.3. 
Again for the assortative and the neutral network, the two 
functions almost overlap, however, for the assortative 
network, the two functions are distinct. \\

\begin{figure*}
\centerline{\epsfxsize=13.25cm \epsfbox{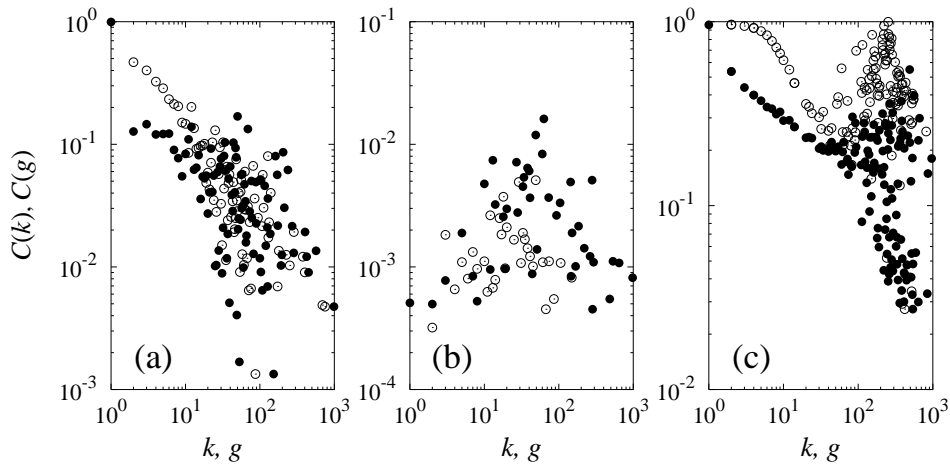}} 
\caption{Plot of the clustering coefficient as a function 
of degree ($\circ$) and BC ($\bullet$) 
for (a) the Internet on the level of autonomous systems (dissortative),
(b) the non-degenerate configuration model with $\gamma=3$ (neutral),
and (c) the coauthorship network in the field of neuroscience (assortative).}  
\label{fig3}
\end{figure*}

To compare the contributions of degree and BC to the 
robustness of networks, we also study the relative size of the giant 
cluster $S$ as a function of the fraction of removed vertices $f$~\cite{cohen}.
We measure $S$ in the two ways of vertex removal 
following the order of (i) degree and (ii) BC for an assortative 
network. 
Fig.~4 shows the data for the coauthorship network 
in the neuroscience field. The relative size $S$ by the vertex removal 
in BC order decreases faster than in degree order up to 
$f \simeq 0.15$, however, for $f > 0.15$, the two data sets
almost overlap. That is because the degree-BC relation 
of Eq.~(1) holds up to roughly $g^* \simeq 8$ and breaks down 
beyond $g^*$, which corresponds to $k^* \simeq 15$. 
The fraction of vertices having degree $k > k^*$ is roughly
$f\simeq 0.15$.
We note that from the point of view of intentional attack,
attack in BC order is more efficient than in degree order.\\

In conclusion, we have examined the BC-BC correlation for 
the three types of scale-free networks, the dissortative, 
the neutral, and the assortative network. While the BC-BC 
correlation behaves similarly to the degree-degree correlation
for the first two types, the BC-BC 
relation is nontrivial for the last type, 
and the mean BC of neighbors 
of a vertex with BC $g_i$ increases with increasing $g_i$ 
but very weakly, being almost independent of $g_i$. 
Such a behavior arises from the fact that the BC of 
the vertex with large degree is not always high, but 
takes rather widely ranged values.\\ 

The work is supported by Grant No. R01-2000-000-00023-0 from the 
BRP program of the KOSEF.\\ 

\begin{figure}[b]
\centerline{\epsfxsize=9cm \epsfbox{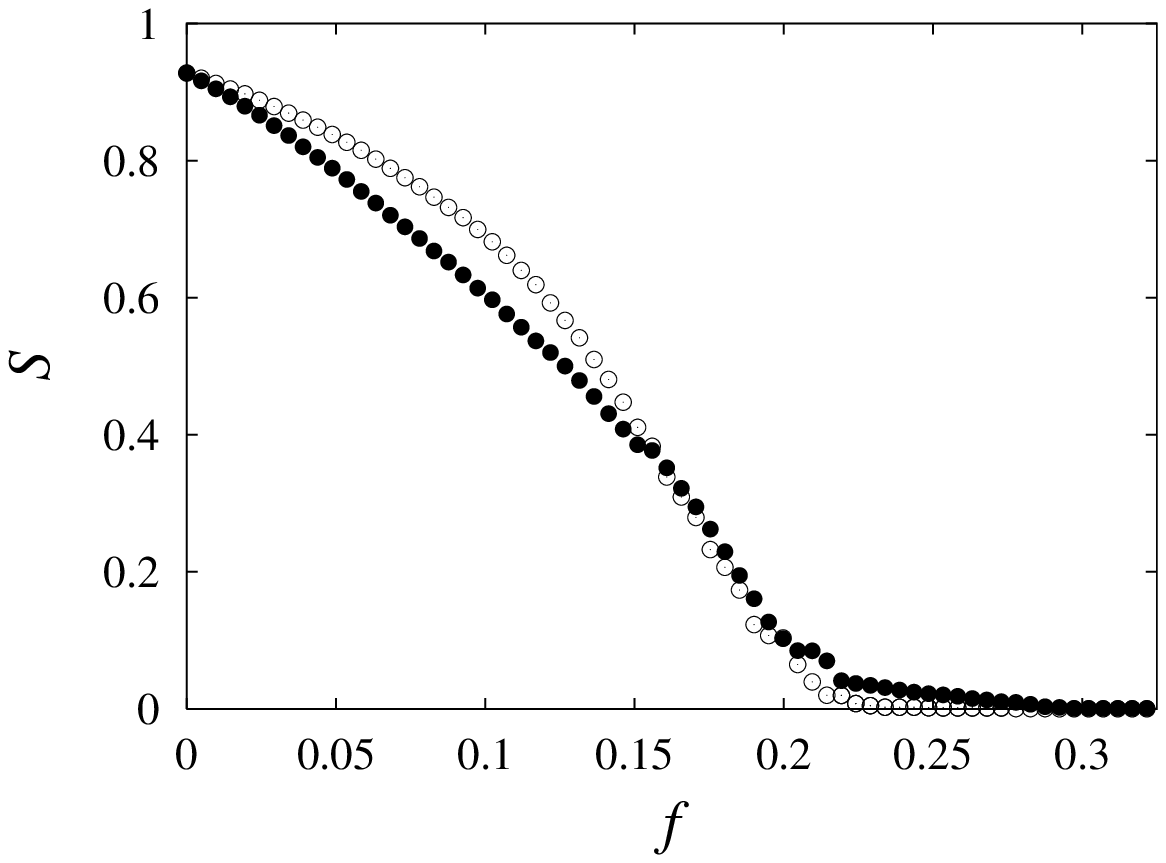}} 
\caption{Plot of the relative size $S$ of the giant cluster 
as a function of the fraction $f$ of removed vertices 
following the order of degree ($\circ$) and BC ($\bullet$) 
for the coauthorship network in the field of neuroscience 
(assortative).}  
\label{fig4}
\end{figure}


\begin{thebibliography}{99}
\bibitem{definition} Given a degree sequence 
${\cal D}\equiv\{k_i\}$ $(i=1,\dots, N)$, we generate a maximally 
random network whose degree sequence is ${\cal D}$, with a constraint 
that any self-loops and multiple edges are forbidden. 
By choosing ${\cal D}=\left\{\left\lfloor
(N/i)^{1/(\gamma-1)}\right\rfloor\right\}$, we get a SF network 
with degree exponent $\gamma$, where $\lfloor \cdot\rfloor$ 
is the floor function.
\end{thebibliography}
\end{document}